\documentclass[a4paper,fleqn,usenatbib,useAMS]{mnras}

\usepackage{graphicx}	
\usepackage{amsmath}	
\usepackage{amssymb}	
\usepackage{multicol}        
\usepackage{bm}		
\usepackage{pdflscape}	

\usepackage[T1]{fontenc}
\usepackage{ae,aecompl}

\usepackage{newtxtext,newtxmath}
\usepackage{enumitem}

\title[Cosmic rays from Galactic starburst activities]{Cosmic rays escaping from Galactic starburst-driven superbubbles}
\author[Z. Zhang, K. Murase and P. M\'{e}sz\'{a}ros]{Zhaowei Zhang$^{1}$\thanks{Contact e-mail: \href{mailto:zzz17@psu.edu}{zzz17@psu.edu}}, Kohta Murase$^{1, 2, 3, 4}$ and Peter M\'{e}sz\'{a}ros$^{1, 2, 3}$ 
\\
$^{1}$Department of Astronomy \& Astrophysics, Pennsylvania State University, University Park, Pennsylvania 16802, USA\\
$^{2}$Department of Physics, Pennsylvania State University, University Park, Pennsylvania 16802, USA\\
$^{3}$Center for Particle and Gravitational Astrophysics, Pennsylvania State University, University Park, Pennsylvania 16802, USA\\
$^{4}$Center for Gravitational Physics, Yukawa Institute for Theoretical Physics, Kyoto, Kyoto 606-8502, Japan}

\date{}

\pubyear{}

\begin{document}
\label{firstpage}
\pagerange{\pageref{firstpage}--\pageref{lastpage}}
\maketitle

\begin{abstract}
We calculate spectra of escaping cosmic rays (CRs) accelerated at shocks produced
by expanding Galactic superbubbles powered by multiple supernovae producing a continuous
energy outflow in star-forming galaxies. We solve the generalized Kompaneets equations 
adapted to expansion in various external density profiles, including exponential
and power-law shapes, and take into account that escaping CRs are dominated by those around 
their maximum energies. 
We find that the escaping CR spectrum largely depends on the specific density profiles and power source properties, 
and the results are compared to and constrained by the observed CR spectrum. 
As a generic demonstration, we apply the scheme to a superbubble occurring in the centre of the Milky Way, and find that under specific parameter sets the CRs produced in our model can explain the observed CR flux and spectrum around the second knee at $10^{17}$ eV. 
\end{abstract}


\section{Introduction}
\label{sec:introduction}

Galaxies whose nuclear regions undergo large star-formation episodes will have substantial rates of
OB star-formation leading to supernovae (SNe) that produce essentially continuous outflows of 
gas and kinetic energy, resulting in superbubbles that shock the ambient interstellar medium (ISM) gas~\citep[e.g.][]{Castor1975, Basu1999}. 
In contrast to superbubbles powered by stellar winds, 
those formed by multiple SNe can be much stronger \citep[for a detailed treatment of stellar 
wind case, see e.g.][]{Castor1975, Olano2009}.  Both protons and 
electrons can be accelerated in such shocks, where high-energy protons can escape the bubble and 
interact with ambient ISM, while electrons can undergo synchrotron and inverse Compton (IC) radiation. 

Powerful superbubbles may be expected in the so-called ultraluminous infrared galaxies (ULIRGs),
defined as galaxies with an infrared luminosity greater than $10^{12} L_{\odot}$ \citep[e.g.][]{Soifer1984}, 
and hyperluminous ones defined as galaxies with an infrared luminosity greater than $10^{13}L_{\odot}$ 
\citep[e.g.][]{Rowan2010}. These ultra- and hyper-luminous galaxies are expected to have large gas densities. Thus, they can host extreme star-forming activities 
\citep[for reviews, see e.g.,][]{Sanders1996}.  \cite{Rowan2017} recently discovered galaxies with 
extreme starburst activities with star formation rates (SFRs) over $5,000~\text{M}_{\odot} 
\text{yr}^{-1}$, with a maximum of $30,000~\text{M}_{\odot} \text{yr}^{-1}$. If a 
normal initial mass function is assumed (e.g.~a Salpeter one), galaxies with such intensive starburst 
activities can produce a large amount of OB stars that would end up as SNe. Provided the rate is high enough and the starburst region is confined in a relatively small and central region of the galaxy, the multiple SNe can be treated as a single energy source with a continuous energy input rate, leading to the formation of a superbubble~\citep[e.g.][]{Anantharamaiah2000}. 

Depending on different density profiles of the ambient gas, the dynamics of the superbubble propagation 
can be very different \citep[for a review, see e.g.,][]{Bisnovatyi1995}. There are only a few  exceptional 
cases for which the analytical solutions describing the bubble propagation can be found. For example, 
Kompaneets' solution \citep[][]{Kom1960} can be used to describe the propagation in an 
exponential-decay density profile along one direction, and \cite{Olano2009} solved the 
power-law density profiles and several other scenarios by expanding the solutions into power series. 
For our purposes and based on the availability of analytic solutions, we list here the density profiles 
that will be discussed in this paper: 
\begin{enumerate}[label=\arabic*)]\itemsep1pt
\item Constant density profile, $\rho(r)\equiv \rho_{0}$,
\item Vertically exponential density profile, $\rho(z) = \rho_{0}\exp{(-|z|/H)}$,
\item Vertical power-law density profile, $\rho(z) = \rho_{0} / (1 + |z|/H)^{\gamma}$ ($\gamma = 1, 2$),
\item Radial power-law density profile, $\rho(r) = \rho_{0} / (1 + r/H)^{\gamma}$ ($\gamma = 1, 2$).
\end{enumerate}

Diffusive shock acceleration (DSA) is generally considered as the main mechanism to accelerate the cosmic
rays (CRs) to high energies \citep[e.g.][]{Krymsky1977, Axford1977, Bell1978, Blandford1987}, which naturally 
produces a $\propto\varepsilon^{-2}$ spectrum of accelerated CRs, provided that the Mach number is 
sufficiently large. 
However, depending on dynamics, the spectrum of the CRs that escape from 
the shock can be different from a simple $\propto\varepsilon^{-2}$ form \citep[e.g.][]{Ohira2010, Drury2011}. In this paper, we assume that the 
CRs are accelerated by the superbubble shock via DSA, and during the CR acceleration process for any given time, low-energy CRs remain confined 
and only the highest-energy CRs can leave the acceleration region from the upstream. Thus, the 
instantaneous spectra of CRs differ from a power-law $\propto\varepsilon^{-2}$. We also take into account the resulting time-dependent effect.


Our calculations are based on combining the Kompaneets' bubble/shock solutions in the presence of different external density profiles, and under the assumption that the accelerated CRs with the highest energies escape from the bubble shell. We study in detail the escaping CR spectra. The advantages of our bubble model are that (a) it is based on simple assumptions, (b) the dynamical evolution requires only the basic physics, (c) the spatial and temporal information of the bubble can be easily retrieved at any epoch, (d) the cumulative escaping CRs spectra (which is focused on in this paper) as well as the escaping CRs spectrum at a single epoch can both be obtained, (e) other accelerated particles, such as electrons, can be studied simultaneously, (f) the calculation can be extended into other applications. 

The paper is structured as follows: in Section $\ref{sec:EnergyInputRate}$, we derive the energy input 
rates powering the superbubbles based on our assumptions; in Section $\ref{sec:DynamicsInDifferentDenisities}$, 
we describe the Kompaneets equations and derive generic solutions for bubbles propagating in different 
ambient gas density profiles; in Section $\ref{sec:CREscapingSpectrum}$, we discuss our assumptions and 
methods to calculate the spectra of escaping CRs, and present our calculated  CR spectra; In Section $\ref{sec:Application}$, we apply our model to a possible superbubble originating from the centre of the Milky Way, 
and discuss the calculated CR flux spectra; in Section $\ref{sec:Summary}$, we summarise the paper.

\section{Energy Input Rate}
\label{sec:EnergyInputRate}

For a starburst galaxy with an extreme star formation rate of $\text{SFR} = \text{SFR}_{4}\times 10^{4}~\text{M}_{\odot}\text{yr}^{-1}$, a Salpeter initial mass function $\psi \propto m^{-\alpha}$, where $\alpha\simeq 2.3$ for $0.08~\text{M}_{\odot} < m < 60~\text{M}_{\odot}$, implies that the formation rate of OB stars is
\begin{equation}
\dot{N}_{\star}(\text{OB}) = \int^{60}_{8}\psi(m)\text{d}m \simeq 78~ \text{SFR}_{4} ~\text{yr}^{-1} .
\end{equation}
The upper and lower limits 0.08 $\text{M}_{\odot}$ and 60 $\text{M}_{\odot}$ are the minimal and maximal stellar masses considered here, below which the hydrogen fusion cannot occur, and above which the radiation pressure is beyond the Eddington limit. The 8~$\text{M}_{\odot}$ is assumed to be the minimal mass of an OB star.

We take the simplifying assumptions that all of these OB stars are born, each resulting in a
typical SN that releases a kinetic energy of $E_{\text{ej}, \text{SN}} = E_{\text{ej}, \text{51}} \times 10^{51}\text{erg}$, all of the SNe happen in a relatively central region of the host galaxy, and the starburst time-scale of this galaxy is longer than the average lifetime of an OB star (millions of years), so that the resulting SNe can be treated as a single event with a constant rate in the centre of the galaxy during the starburst time scale. The energy input rate from such an energy source is:
\begin{equation}
\begin{split}
L_{0}\equiv\dot{E_{0}}  &= \dot{N}_{\star}(\text{OB})\cdot E_{\text{ej},\text{SN}} \\
&\simeq2.5\times 10^{45} \cdot \text{SRF}_{4} \cdot E_{\text{ej}, \text{51}}~\text{erg}~ \text{s}^{-1}.
\end{split}
\label{eq:EneInRateConstant}
\end{equation}

This constant and continuous energy source is assumed to decay rapidly after a specific starburst activity time-scale $\tau$, 
\begin{equation}
L(t) = L_{0}\cdot\exp{(-t / \tau)}
\label{eq:EneInRateExponentialDecay}~,
\end{equation}
where $\tau$ can be in order of $\sim10$~Myr \citep[e.g.][]{Mannucci2010, Tacconi2013}.

If such a source is present in the centre of a galaxy, the continuous input of kinetic energy would produce a shock that accelerates the ambient gas, which leads to the formation of an outgoing shell, hence a bubble is formed. The mass injected from the source can be ignored compared with the mass swept by the shock, and the kinetic energy carried by the injection would quickly dissipate into the thermal energy of the bubble \citep[][]{Castor1975}. The propagation of the bubble shell can be described analytically in a few limited cases, and we will present and derive the analytical solutions for several specific scenarios in the following section.

\section{dynamics of a superbubble expanding in different ambient gas density profiles}
\label{sec:DynamicsInDifferentDenisities}

\subsection{Constant Density Profile}

We follow the procedure presented by \cite{Castor1975}, who derived the detailed solutions of bubble dynamics in a homogeneous density profile, assuming a constant energy input source in the bubble centre. The first phase of the bubble propagation is the adiabatic phase (a large, hot, low-density bubble interior with a thick, hot, dense, swept ISM shell), when the shell temperature is high and radiation cooling is not sufficient. During this phase, the shell radius can be written as a function of time:
\begin{equation}
R_{\text{s}}(t) = 0.88\cdot\left(\frac{L_{0}}{\rho_{0}}\right)^{1/5}t^{3/5},
\label{eq:RsConstantDensityProfileAdiabeticPhase}
\end{equation}
where 0.88 is a constant obtained by similarity solutions. The time dependence here is $t^{3/5}$ instead of $t^{2/5}$ as that for the supernova remnant (SNR) scenario, because of an extra time dependence in $L_{0}$ compared with a single $E_{0}$ as that in the SNR. 

Supposing that the ambient gas density is a constant with $\rho_{0} = \rho_{0,-21}\times 10^{-21}\text{g}\cdot \text{cm}^{-3}$, then with Eq. ($\ref{eq:EneInRateConstant}$),
\begin{equation}
R_{\text{s}}(t) \simeq 2.1\times 10^{21}\cdot\left(\frac{\text{SFR}_{4}\cdot E_{\text{ej}, 51}}{\rho_{0, -21}}\right)^{1/5}t_{\text{Myr}}^{3/5} ~\text{cm}~,
\label{eq:RsConstant}
\end{equation}
where $t = t_{\text{Myr}}\times 1 \text{Myr}$ and the velocity can be obtained by taking time derivative of the radius,
\begin{equation}
v_{\text{s}}(t) \simeq 6.8\times 10^{7}\cdot\left(\frac{\text{SFR}_{4}\cdot E_{\text{ej}, 51}}{\rho_{0, -21}}\right)^{1/5}t_{\text{Myr}}^{-2/5}~\text{cm}~ \text{s}^{-1}.
\end{equation}
During this phase, the interior of the bubble contains 5/11 of the totally released kinetic energy and the shell contains the rest 6/11, in terms of both kinetic and thermal energies. 
(We discuss how this value changes for other density profiles.)

If we assume the bubble shell transfers $\epsilon_{B}$ fraction of its kinetic energy into magnetic energy, then
\begin{equation}
\frac{B^{2}}{8\pi} = \epsilon_{B}\cdot \frac{1}{2}\rho_{0}v_{\text{s}}^{2}~,
\label{eq:BFieldKineticE}
\end{equation}
the maximum energy of the ions that can be accelerated by the shell is \citep[][]{Drury1983} 
\begin{equation}
\begin{split}
\varepsilon_{\text{max}}(t) &\simeq \frac{3}{20}\cdot Z\cdot e\cdot B\cdot R_{s} \cdot \frac{V_{\text{s}}}{c} \\
&\simeq1.6\times 10^{17}~ Z~\epsilon_{B, -2}~(\text{SFR}_{4}\cdot E_{\text{ej}, 51})^{3/5}\rho_{0, -21}^{-1/10}~ t_{\text{Myr}}^{-1/5}~\text{eV}~,
\end{split}
\label{eq:EMax}
\end{equation}
where $\epsilon_{B, -2} = \epsilon_{B}/0.01$ and $Z$ is the atomic number of ions. Thus, from this simple analysis we demonstrate that the maximum energy of an ion that can be accelerated by such a bubble is above 100 PeV, assuming characteristic parameters are set to 1. 

One thing worth noticing is that $\epsilon_{B}$ here has considerable uncertainties, which should be studied in a case-by-case basis. If both the upstream and downstream magnetic fields at present are known for a specific case, then $\epsilon_{B}$ can be obtained by normalising the calculations to the known value at the current epoch, then the efficiency can be fixed and applied to earlier epochs. But for simplicity and generality, $\epsilon_{B} = 0.01$ is used in this and following sections.

For the SNR, since they lack any other kinetic energy input after the initial explosion, the conservation of momentum can be applied to calculate the dynamics after the bubble leaving the adiabatic phase, when the thick shell of the bubble begins to cool down and is compressed into a thin one \citep[][]{Zeldovich1966}, `snowplowing'  the newly swept ISM gas. For a SN-driven superbubble, however, there is still a constant energy input after the shell leaves the adiabatic phase, provided the source lasts long enough. Thus, except that the thick shell has been compressed into a thin and cold one, the dynamics is similar to that in the adiabatic phase: 
\begin{equation}
R_{\text{s}}(t) = 0.76\cdot\left(\frac{L_{0}}{\rho_{0}}\right)^{1/5}t^{3/5} .
\label{eq:RsConstantDensityProfileSnowPlowPhase}
\end{equation}
The only difference between Eq.~($\ref{eq:RsConstantDensityProfileSnowPlowPhase}$) and Eq.~($\ref{eq:RsConstantDensityProfileAdiabeticPhase}$) is the prefactor, which changes from 0.88  into 0.76. The discontinuity between the equations can be interpreted as the compressing of the bubble shell. In this paper, we use the Eq.~($\ref{eq:RsConstantDensityProfileAdiabeticPhase}$) as the reference for the bubble propagating in a constant density profile, since it is consistent with the assumptions and calculations of the Kompaneets approach to the dynamics of shocks propagating under various density profiles.



In the above calculations we have assumed that the energy input is constant in time. More realistically, the 
energy input may decline gradually towards later epochs; however, such a drop of the energy input does not have a 
substantial effect on the solutions for the constant density profile case. For density profiles which vary, however, 
the solutions can break down \citep[][]{Basu1999}. We discuss this in some more detail in Section 
$\ref{sec:ThermalEnergyIssue}$.

\subsection{Vertically Exponential Decay: $\exp{(-|z|/H)}$}

This is the scenario that Kompaneets initially considered \citep[][]{Kom1960}. In this case $\rho(z) = \rho_{0}\exp{(-|z|/H)} = \rho_{0}F(z)$, where $\rho_{0}$ is the central plane gas density and $H$ is a characteristic scale height (for example, the scale height of the Galactic disc). There are three fundamental assumptions in Kompaneets' solutions: a), the pressure inside the bubble is uniform (isobaric) and dominant over the external pressure; b), the mass swept up by the shock is trapped in a thick shell following the shock; c), each element of the shell is moving along the direction of the force behind it (e.g. each element is moving vertically to the tangent plane cutting the element on the shell). Thus the evolution of a shock (shell) front generated by a point energy source can be represented by a function $f(x, y, z\cdots ; t) = 0$ \citep[][]{Kom1960}. Since the density decreases along $z$ (and negative $z$) direction, it's easier to describe the dynamics in a cylindrical system of coordinate $(r, z)$, where $z$ is perpendicular to the stratification plane. The dependence on azimuthal angle can be ignored because the solution is symmetric around the $z$-axis. Therefore the evolution of a shock front generated by a point explosion can be described by a function $f(r, z, t) = 0$ \citep[][]{Olano2009}. 

At the shock (shell) front $\text{d}f/\text{d}t = 0$, hence $(\partial f/\partial r)(\text{d}r/\text{d}t) + (\partial f/\partial z)/(\text{d}z/\text{d}t) + \partial f/\partial t = \mathbfit{v}\cdot\nabla f + \partial{f}/\partial{t} = 0$, and based on the assumption that $\mathbfit{v}$ and $\nabla f$ are parallel vectors,
\begin{equation}
v = |\mathbfit{v}| = -\frac{\partial f/\partial t}{|\nabla f|}.
\end{equation}

By further assuming that the equation $f(r, z, t) = 0$ has a solution such that $r$ depends on $z$ and $t$ explicitly, then $r = g(z, t)$ and $f(r, z, t) = r - g(z, t) = 0$. Thus $\partial f/\partial z = - \partial g/\partial z = - \partial z/\partial z$ and $\partial f/\partial r = 1$, and
\begin{equation}
|\nabla f| = \sqrt{\left(\frac{\partial f}{\partial r}\right)^{2} + \left(\frac{\partial f}{\partial z}\right)^{2}} = \sqrt{1 + \left(\frac{\partial r}{\partial z}\right)^{2}}~,
\end{equation}
which leads to
\begin{equation}
\left(\frac{\partial r}{\partial t}\right)^{2} - v^{2}\left[1 + \left(\frac{\partial r}{\partial z}\right)^{2}\right] = 0~.
\label{eq:KomDifferentialEquationToSolve}
\end{equation}

Since the internal pressure is assumed to be uniform and dominant over the external pressure, the velocity of the shock can be obtained using the strong shock conditions
\begin{equation}
v = \sqrt{\frac{\gamma_{\text{ad}} + 1}{2}\frac{P(t)}{\rho(z)}}
\label{eq:VolocityRamPressure}
\end{equation}
and the pressure is related to the thermal energy as
\begin{equation}
P(t) = (\gamma_{\text{ad}} - 1)\frac{E_{\text{th}}}{V(t)}~,
\label{eq:Pressure}
\end{equation}
where $\gamma_{\text{ad}}$ is the adiabatic index (we use 5/3 in this paper) and $E_{\text{th}}$ is the thermal energy inside the bubble, and $V(t)$ is the bubble volume. Thus the velocity of the shock (shell) front can be written as
\begin{equation}
v^{2} = \frac{E_{\text{th}}(\gamma_{\text{ad}}^{2} - 1)}{2\rho_{0}V(t)}F(z)^{-1}~,
\label{eq:v2KomOrign}
\end{equation}
which can be then inserted into Eq.~($\ref{eq:KomDifferentialEquationToSolve}$) to solve $r$. 

However, it is hard, if not impossible, to solve Eq.~($\ref{eq:KomDifferentialEquationToSolve}$) explicitly, thus Kompaneets used an intermediate factor $y$ to solve the equation \citep[e.g.][]{Kom1960}:
\begin{equation}
y = \int^{t}_{0}\left(\frac{E_{\text{th}} (\gamma_{\text{ad}}^{2} - 1)}{2 \rho_{0}V(t)}\right)^{1/2}dt
\label{eq:Y}
\end{equation}
with the help of which
\begin{equation}
\left(\frac{\partial r}{\partial y}\right)^{2} - F(z)^{-1}\left[1 + \left(\frac{\partial r}{\partial z}\right)^{2}\right] = 0~,
\end{equation}
which has a solution:
\begin{equation}
\begin{split}
r(z, y) = ~ & 2H \\
&\times \arccos{\left[\frac{1}{2}\exp{\left(\frac{z}{2H}\right)}\left(1-\frac{y^{2}}{4H^{2}} + \exp{\left(-\frac{z}{H}\right)}\right)\right]}.
\label{eq:KomSolu}
\end{split}
\end{equation}

Eq.~($\ref{eq:KomSolu}$) depends explicitly on $z$, the vertical component in cylindrical coordinate; and $y$, a factor that includes all other information: thermal energy in the interior bubble $E_{\text{th}}$, volume of the bubble $V(t)$, central density $\rho_{0}$, adiabatic index $\gamma_{\text{ad}}$, and time $t$. By combining the equality that $\tan{\theta} \equiv z / r$, $r$ can be solved at each horizontal angle $\theta$ at each `time' $y$. The solution has a range $y \in [0, 2H]$ such that above this the solution no longer holds and the bubble quickly expands to infinity, the so-called `break out' limit (the literature sometimes defines `break out' differently, but in this paper we consider that a bubble `breaks out' when an analytic solution no longer holds, or the solution begins to lead to an explosive behaviour of propagation along any direction, or some physical quantities become no longer conserved, for example, the total energy). 

\subsection{Radial Power Law: $1/(1+r/H)$}

In this case $\rho(r) = \rho_{0}/(1+r/H) = \rho_{0}F(r)$ that leads to a spherically symmetric solution. Thus the dynamics of the shock front can be simply written as $(\text{d} r/\text{d} t)^{2} = v^{2}$, where $r$ is in spherical coordinate. By introducing the same intermediate factor $y$,  the dynamics in terms of $y$ is:
\begin{equation}
\left(\frac{\partial r}{\partial y}\right)^{2} -F(r)^{-1} = \left(\frac{\partial r}{\partial y}\right)^{2} - (1 + r/H) = 0~,
\label{eq:ComInPLR1}
\end{equation}
which has a solution in a simple form:
\begin{equation}
r(y) = \frac{y(4H + y)}{4H}.
\label{eq:PL1RSolution}
\end{equation}

Differently from the previous case, here there is no limit on $y$. At large $H$ ($\sim$ constant density), Eq.~($\ref{eq:PL1RSolution}$) reduces to $r(y)\simeq y$, which, combined with Eq.~($\ref{eq:Y}$) and the assumption that the energy injection is constant, shows that $r$ converges back to Eq.~($\ref{eq:RsConstantDensityProfileAdiabeticPhase}$), confirming the consistency between Kompaneets' approach and the approach by \cite{Castor1975}.  

\subsection{Radial Power Law: $1/(1+r/H)^{2}$}
\label{sec:RadialPowerLaw2}

This is similar to the previous section, the only difference is that Eq.~($\ref{eq:ComInPLR1}$) now changes into:
\begin{equation}
\left(\frac{\partial r}{\partial y}\right)^{2} -F(r)^{-1} = \left(\frac{\partial r}{\partial y}\right)^{2} - (1 + r/H)^{2} = 0
\end{equation}
with a solution
\begin{equation}
r(y) = H[\exp{(y/H)} - 1]~.
\label{eq:PL2RSolution}
\end{equation}

\subsection{Vertical Power Law: $1/(1+|z|/H)$}

In this case $\rho(z) = \rho_{0} / (1+|z|/H)$, and \cite{Olano2009} expanded the solution into a series of powers to solve this problem. Here we quote their results:
\begin{equation}
\sqrt{r^{2} + z^{2}} = 2Ht_{\star} + Ht_{\star}^{2}\sin{\theta}~,
\label{eq:PL1ZSolution}
\end{equation}
where $t_{\star} = y / 2H$, $\theta$ is the horizontal angle between a point on the shell and horizontal plane, and $r$ and $z$ are in cylindrical coordinates. This equation can be solved at each $t_{\star}$ with $\sin{\theta} \equiv z/\sqrt{r^{2} + z^{2}}$. Graphically Eq.~($\ref{eq:PL1ZSolution}$) represents a cardioid shape like a bubble (upside down).

\subsection{Vertical Power Law: $1/(1+|z|/H)^{2}$}
Also following \cite{Olano2009} the solution to this density profile is:
\begin{equation}
r^{2} + (z - 2H\sinh^{2}{t_{\star}})^{2} = (H\sinh{2t_{\star}})^{2}.
\label{eq:PL2ZSolution}
\end{equation}
This shape is also easy to visualize: it is a circle (shell in 3-D) with radius $H\sinh{2t_{\star}}$ centred at $(r, z) = (0, 2H\sinh^{2}{t_{\star}})$, where $r$ and $z$ are in cylindrical coordinate. 
\\
\\
To summarise, Eqs.~($\ref{eq:KomSolu}$), ($\ref{eq:PL1RSolution}$), ($\ref{eq:PL2RSolution}$), ($\ref{eq:PL1ZSolution}$), and ($\ref{eq:PL2ZSolution}$) are the solutions for the bubble dynamics in different density profiles, the implications of which will be discussed in later calculations.  Simple evaluations demonstrate that these solutions represent similar dynamics at small values of $y'$ (e.g. at early times). This is reasonable because all densities can be approximately taken constant at early stages, hence a constant density profile solution is expected in all cases. However, they can be very different at later epochs: solutions for density profiles decreasing along $z$ directions result in elongated-shape bubbles, while those decreasing along the $r$ direction remain spherical ones. Different solutions also lead to different speeds of the bubble propagation depending on how fast the density drops. For example, a power law with $\gamma = 2$ results in a bubble that propagates faster than power laws with $\gamma = 1$ assuming the power of the sources and the central gas densities are identical, and the exponential density profile leads to a bubble that propagates much faster along the direction of the density decaying than in a constant case. 

\section{Escaping CR Spectra for Different Density Profiles}
\label{sec:CREscapingSpectrum}

As the bubble propagates forwards, the shock sweeps up and accelerates the ambient gas, and the maximal energy of the CRs accelerated in the shock can be approximated via Eq.~($\ref{eq:EMax}$). While the majority of the accelerated protons will be trapped in the shell, some can escape from it \citep[e.g.,][]{Caprioli2010, Ohira2010, Drury2011}. 
These escaped protons can of course further interact with the ambient gas outside, and undergo other interactions. For example, the resulting $pp$ interactions will produce high energy neutrinos and gamma-rays \citep[e.g.][]{Senno2015, Xiao2016}. Thus, the spectra of escaping CRs under different density profiles are of interest to study. 

CRs with maximally accelerated energies are assumed to escape from the bubble shell at each epoch, which is a specific application of the `escape-limited' model by \cite{Ohira2010}. As we shall discuss later, we apply this model in an approximate manner. The spectrum of escaping CRs is calculated through the following steps:

\begin{enumerate}[label=(\alph*)]
\item We write the solutions of the shock (shell) front in terms of horizontal angles $\theta$ at $y$, e.g. $r(\theta, y)$,
\item We calculate the corresponding time at each $y$,
\item We `cut' the shell into 10 equal pieces with equal surface areas at each epoch, and obtain the corresponding horizontal angles that divide the equal surface areas,
\item We assume a certain fraction of the kinetic energy of the shell is used to accelerate CRs, and that CRs with the maximally accelerated energies, approximated by Eq. ($\ref{eq:EMax}$), escape the system,
\item We integrate the amount of CRs over the areas defined by step (c) to obtain the CR escaping flux at a certain epoch at a certain angle:
\end{enumerate} 
\begin{equation}
\centering
\varepsilon\frac{\text{d}n}{\text{d}\varepsilon \text{d}t} = \int^{\theta_{+}}_{\theta_{-}}\frac{\epsilon_{p, \text{esc}}\rho(\theta, y)v(\theta, y)^{3}}{\varepsilon_{\text{max}}(\theta, y)}\text{d}A~,
\label{eq:CREscapingFlux}
\end{equation}
where $\epsilon_{p, \text{esc}}$ is the fraction of kinetic energy of the shell at certain epoch $y$ and certain angle $\theta$ that is converted into escaping CRs, and those CRs escape with the maximum energy $\varepsilon_{\text{max}}$, which also depends on time and angle. 

For a conventional SN, the fraction of kinetic energy of a shock transferred to accelerated CRs is about $\epsilon_{p}=0.1$, but the energy carried by escaping CRs at a given time is smaller~\citep[e.g.,][]{Ohira2010, Drury2011}. We assume that the accelerated CRs follow a $\text{d}n/\text{d}\varepsilon\propto \varepsilon^{-s_{p}}$ spectrum that can be uniquely determined with $\varepsilon_{\text{min}}$ (that is set to $m_{p}c^{2}$), $\varepsilon_{\text{max}}$ (that is evaluated at each epoch), and the energy fraction $\epsilon_{p}$. 
We here note that the power-law distribution of the accelerated CRs is justified in momentum space. However, the total energy carried by CRs is dominated by the CRs with $\sim1-10~\text{GeV}$ for $s_p\gtrsim2$, and we may approximate the CR spectrum to be a power law in energy space. Then the escaping efficiency is calculated by dividing $\text{d}n/\text{d}\varepsilon|_{\varepsilon_{\text{max}}}\times\Delta \varepsilon\times \varepsilon_{\text{max}}$ over the total accelerated CR energy at a specific epoch and angle. The `width' $\Delta \varepsilon$ is normalised according to the `escape-limited' model by \cite{Ohira2010}, who showed that the escaping CR spectrum follows a `bell' shape that centres around the maximal energy and drops quickly away from the centre. In our model, we assume that CRs accelerated to maximal energies escape from the system, which is equivalent to the statement that the escaping CR energy spectrum forms a `box' shape. The width of the `box', $\Delta \varepsilon$, is calculated such that the escaping energy given by the box is the same as that of the `bell', which turned out to be $\sim 2.28~\varepsilon_{\text{max}}$, free of other parameters. We have verified that different assumptions of the shapes do not lead to  noticeable changes to our results, thus we keep using the `box' shape escaping spectrum in this paper. The benefit of the `box' shape assumption is that we don't need to track each escaping CR spectrum at each epoch at each angle (as in the `bell' shape model), which can save a large amount of computing efforts and time. In this section, a $\varepsilon^{-s_{p}}$ spectrum of accelerated CRs with $s_{p} = 2$ is used. But as we shall discuss in Section $\ref{sec:Application}$, the index $s_{p}$ might not necessarily be 2, and 2.4 is actually possible, which leads to a smaller escaping efficiency, provided the acceleration efficiency does not change. Numerical simulations have shown that $\epsilon_{p}$ can vary \citep[e.g.,][]{Caprioli2014} and spectral indices inferred by observations are steeper \citep[e.g.,][]{MF19}, but in this section we fix $s_{p}=2$ and $\epsilon_{p} = 10\%$ for the generic discussions.

The total CR escaping spectrum can be obtained by integrating Eq.~($\ref{eq:CREscapingFlux}$) over time. However, the direct integration is impossible because of the `cutting' procedure presented earlier, thus instead we multiply the values obtained from Eq.~($\ref{eq:CREscapingFlux}$) by the time interval between two epochs. After binning the total energy range into several energy intervals, the values in each energy intervals are summed to obtain the CR escaping spectrum. Symbolically, the above numerical procedure can be written as:
\begin{equation}
\begin{split}
\varepsilon^{2}\frac{\text{d}N}{\text{d}\varepsilon} =&~ \varepsilon\sum_{\Delta \varepsilon} \Bigg[\sum_{\text{time}}\Delta t\sum_{\text{shell}} \Big( \\
&\int^{\theta_{+}}_{\theta_{-}} \frac{\epsilon_{p, \text{esc}}\rho(\theta, y)v(\theta, y)^{3}}{\varepsilon_{\text{max}}(\theta, y)}\text{d}A \Big) \Bigg]\Bigg|_{\varepsilon_{\text{max}}\in\Delta \varepsilon}~.
\end{split}
\label{eq:CREscapingSpectrum}
\end{equation}
We have tested this calculation scheme against a Sedov-Taylor scenario, which is `simulated' by fixing the initial total energy (to simulate the single explosion) and setting the scale height of an exponential density profile as a very large number (to simulate the constant density). By doing so, the code and scheme remain unchanged, otherwise, a `special treatment' would not meet the purpose of testing the generic calculation scheme. The result is a flat ($\varepsilon^{2}\text{d}N/\text{d}\varepsilon\propto \varepsilon^{0}$) escaping CR spectrum, which is expected from a Sedov-Taylor case \citep[see, e.g.,][]{Ohira2010}.

\subsection{Remarks on Thermal Energy inside a Bubble}
\label{sec:ThermalEnergyIssue}

It is critical to determine the thermal energy in a bubble ($E_{\rm th}$) as it directly affects the dynamics of the propagation. For a constant density profile, it can be safely assumed that the thermal energy inside the bubble is a constant fraction, 5/11, of total input energy \cite[][]{Castor1975}. Thus for the constant density profile, the `thermal luminosity' can be written as 
\begin{equation}
L_{\text{th}}(t) = \frac{5}{11} L(t) = \frac{5}{11}L_{0}\exp{(-t/\tau)} ~.
\label{eq:ConstantDensityTHLuminosity}
\end{equation} 

However, for other cases, a simple 5/11 relationship does  not always hold, as pointed out by \cite{Basu1999}, who found that in the exponential density profile the thermal energy drops very fast below the 5/11 of total energy at later times. This decrease is due to the rapid expansion of the bubble during later phases because of the rapid decline of the external gas densities, hence large amount of interior thermal energy needs to be converted into kinetic energy of the shell to support the rapid expansion. Thus, in this paper, we obtain $L_{\text{th}}(t)$ by fitting it to the numerical results via solving the original formalism of Kompaneets' equations using a Runge-Kutta method (See Appendix A of \cite{Basu1999} for detail). We found that in all non-constant cases, the thermal luminosity drops below the 5/11 of the total input energy very quickly at later times, consistent with the arguments by \cite{Basu1999}. After fitting to the thermal luminosity, we can insert it back into Eq.~($\ref{eq:Y}$) and obtain the time $t$ at a specific $y$. Thus, in later sections, we always make a fit for the thermal luminosity before calculating the CR escaping spectra for non-constant cases.

\subsection{Escaping CR Spectra for Constant and Vertically Exponential Decay Density Profiles}
\begin{figure}
    \centering
    \includegraphics[width=0.45\textwidth]{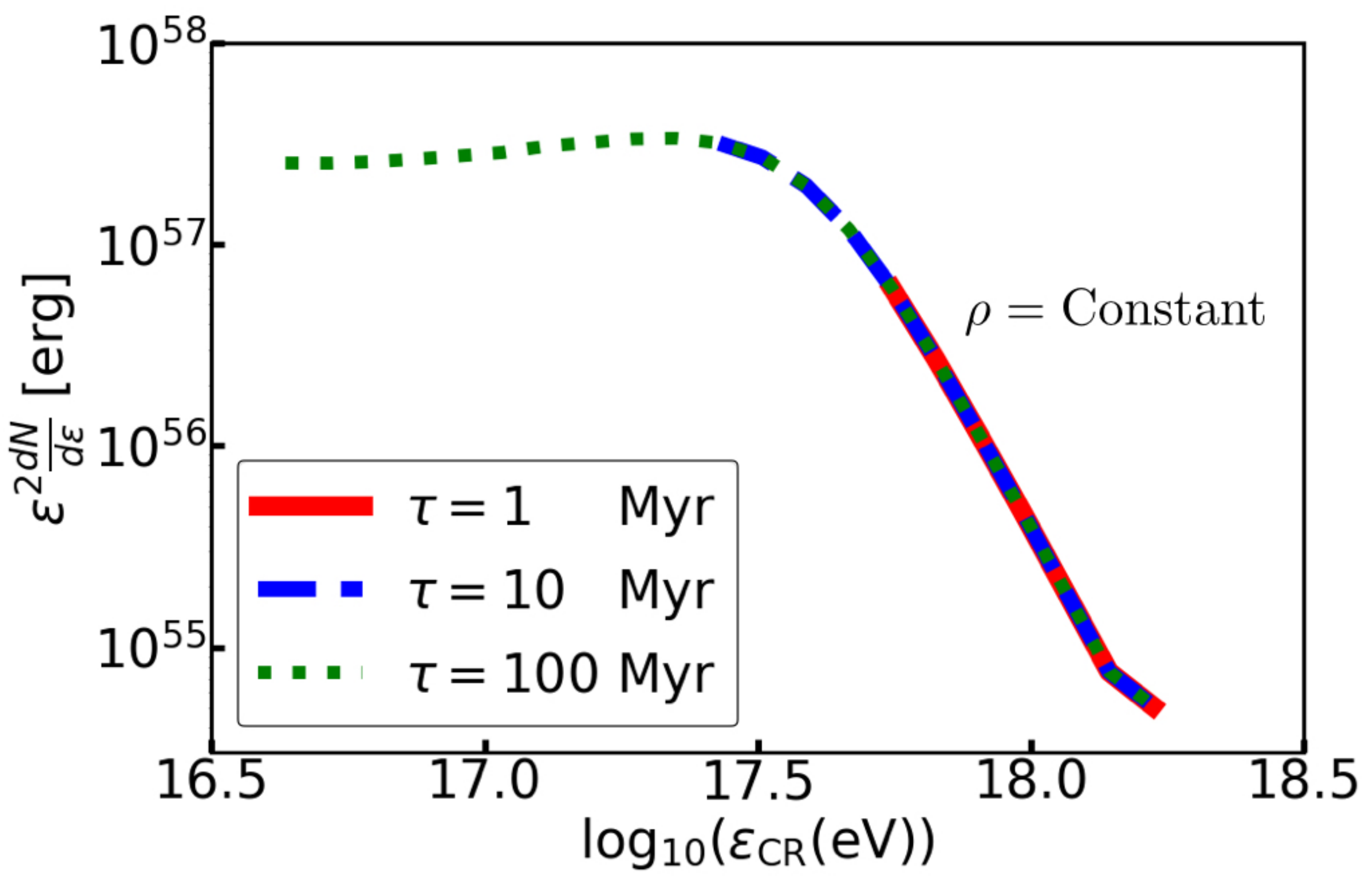}
    \includegraphics[width=0.45\textwidth]{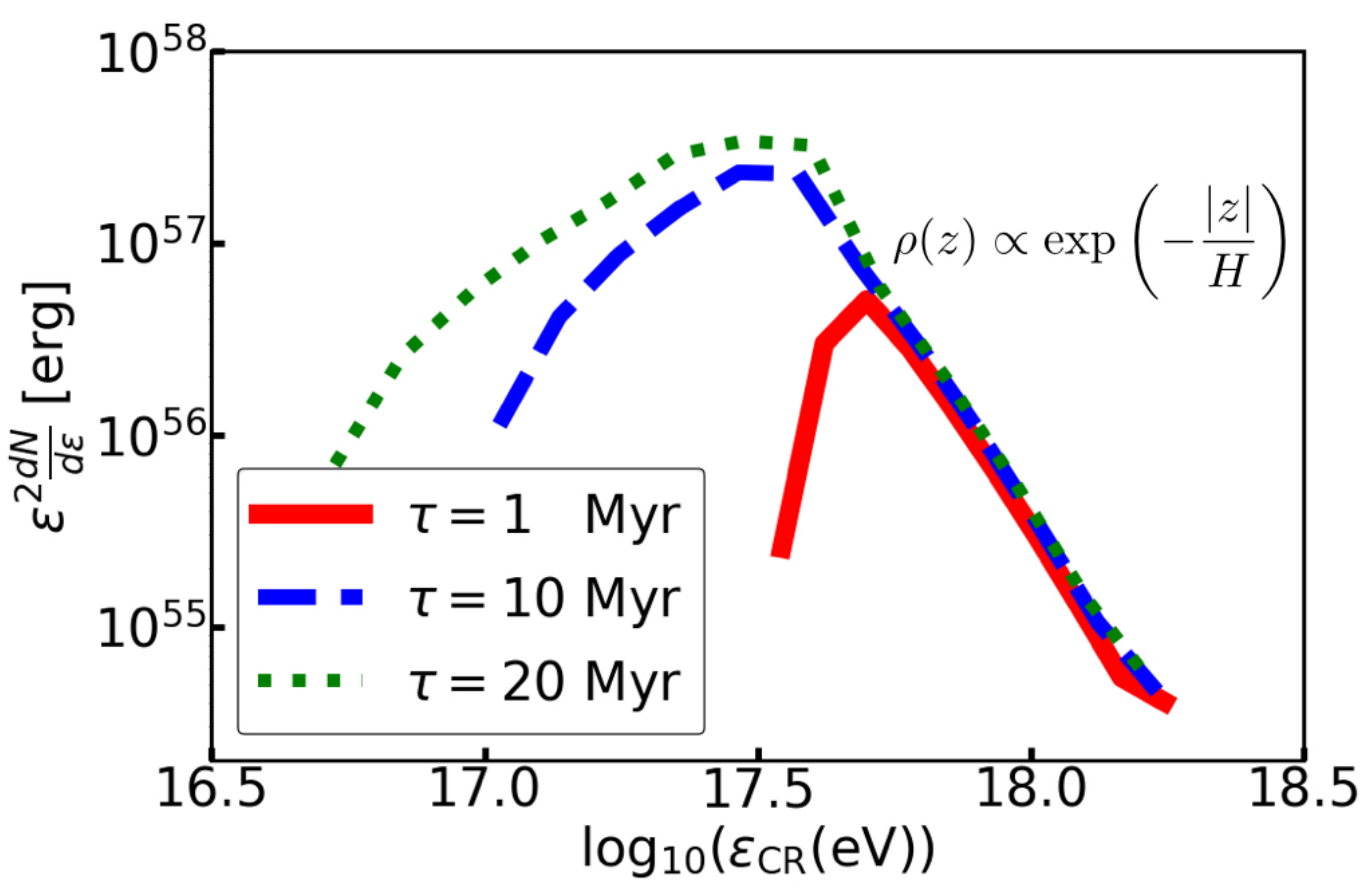}
    \caption{Top: escaping CR spectra from a constant density profile with continuous energy input that stops 
at 10 Myr, accumulated up to different integration times. Bottom: similar plot but in an exponentially-decay density 
profile.}
    \label{fig:1}
\end{figure}

We are interested in the constant and exponentially decaying density profiles because for a galaxy with large SFR, the galactic gas tends to concentrate near the central region, often in the form of a disc \citep[e.g.][]{Sanders1996, Rowan2009}. Thus constant or exponentially decaying density profiles are common descriptions for such scenarios. Differently from the single explosion case, the energy input is continuous in superbubbles with an assumed cutoff at $\tau \simeq 10~\text{Myr}$ \citep[][]{Tacconi2013}. To demonstrate and study the general properties of escaping CR spectra from constant and exponential density profiles, we perform calculations using a specific set of parameters. In real applications, the parameters might vary significantly, but the overall spectral shape and properties resemble the examples here. For the demonstrations we use $\text{SFR}_{4} = 1, E_{\text{ej}, 51} = 1$, and $\rho_{0, 21} = 1$ for both cases, and $H = 1~\text{kpc}$ for the exponential density profile. The results are shown in Fig.~($\text{\ref{fig:1}}$). 

Lines with different colors in Fig.~($\ref{fig:1}$) are CR cumulative escaping spectra up to different time-scales. 
At all epochs, only the newly accelerated CRs with the highest energies escape from the system, 
while the lower-energy CRs remain confined. Following the prescription described above, we evaluate the escaping efficiency 
$\epsilon_{p, \text{esc}}$ in Eq. (26), and find that this efficiency almost stays constant during most of the lifetime of the bubble, and varying only a little at later times. 
Thus, the time-integrated spectrum of the escaping CRs becomes a power law at high energies, because of CRs that escape during the starburst activity. 
Then it becomes saturated because the low-energy component originates from the CRs that escape after the starburst ends.
As indicated by the top plot (constant density), at early times, the escaping CRs produced by the bubble form a spectrum $\text{d}N/\text{d}\varepsilon \propto \varepsilon^{-s_{\text{esc}}}$ where $s_{p, \text{esc}}$ is approximated as $\simeq 6.89$, much steeper than the flat ($s_{p, \text{esc}}=2$) spectrum in the Sedov-Taylor scenario; while at later epochs the spectrum converges to a flat one. 
For the constant density profile at early epochs, Eq.~($\ref{eq:CREscapingSpectrum}$) can be approximated by $\varepsilon^{2}\text{d}N/\text{d}\varepsilon\propto \varepsilon\times t \times v^{3} \times R^{2}/ \varepsilon \propto t \propto \varepsilon^{5}$ (from Eq.~($\ref{eq:EMax}$)), thus $\text{d}N/\text{d}\varepsilon\propto \varepsilon^{-7}$, consistent with numerical results. The small discrepancy arises from the numerical procedures. More generally, it is given by
\begin{equation}
s_{\rm esc}=s_{\rm acc}+\frac{\beta}{\alpha}~,
\end{equation}
where $\varepsilon_{\rm max}\propto R^{-\alpha}$ and $K$, the normalisation factor, $\propto R^\beta$ \citep[][]{Ohira2010}.
The overall shape is reasonable because the energy input is stopped at $\tau = 10$ Myr, after which the system can be treated as a single explosion that happened at some early epoch, thus the solution converges back to a Sedov-Taylor scenario. When considering the spectrum at later times, a transition from non-Sedov-Taylor to Sedov-Taylor situation can be seen. The change of energy dependence can be explained by treating the continuous energy input as the combination of a series of Sedov-Taylor cases that happen continuously. Thus combined, the overall spectrum is very steep at early times but converges to a flat one after the energy injection stops.  

Such behaviour is not apparent for the exponential density profile. The spectrum is identical to that of the constant one at high energies at earlier epochs, which is reasonable because densities are not very different for both cases at early times, and similar argument can be used to explain the steep spectrum.  However, the spectrum behaves differently at lower energies and later times. As indicated by the plot, the escaping CR spectrum peaks at specific energies that depend on the different time-scales up to which they have accumulated. For example, the spectrum accumulated up to 1 Myr peaks at $\sim 700$ PeV, while for the spectrum accumulated at 10~Myr the peak is at $\sim$ 300 PeV. The plot also indicates that the spectra become broader towards lower energies as they get accumulated to larger time-scales.  

One general criticism about Kompaneets' solutions and assumptions is that the solution results in a `break out' behaviour at large horizontal angles at later times, while it is argued that the bubble shell should not be able to follow the shock since the energy is not enough. This is due to the breakdown of one of the Kompaneets' assumptions that the bubble interior pressure is homogeneous and dominant over the external pressure. This condition holds when the interior sound speed is greater than the propagation speed of the shell. Thus when the propagation speed, mathematically determined by the solutions, exceeds the interior sound speed, the assumption no longer holds, and the solutions break down. A forecasting proxy of the `break out' is the rapid decrease of the interior thermal energy, as discussed in Section 4.2. This phenomenon is most obvious for the fast decaying density profiles at later epochs, which leads to an energy conservation problem even when fitting the thermal luminosity with the Runge-Kutta method. Thus the energy conservation always needs to be monitored. This is the reason why the calculations are stopped after $\tau=20$ Myr for the exponential example, after which the energy conservation is no longer held. Thus in this and later sections, we always monitor the overall energy conservations when performing calculations.

\begin{figure*}
    \centering
    \includegraphics[width=0.45\textwidth]{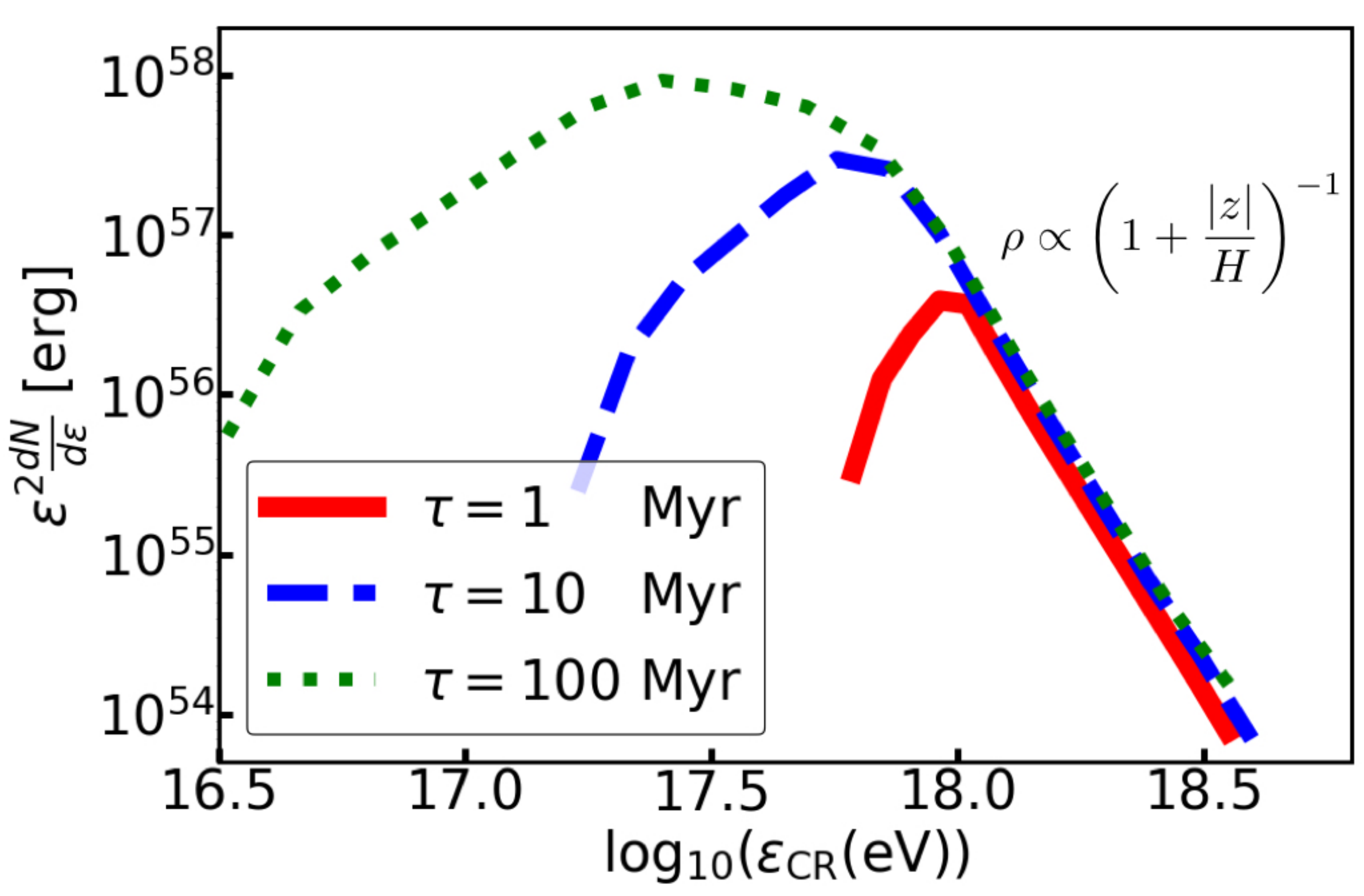}
    \includegraphics[width=0.45\textwidth]{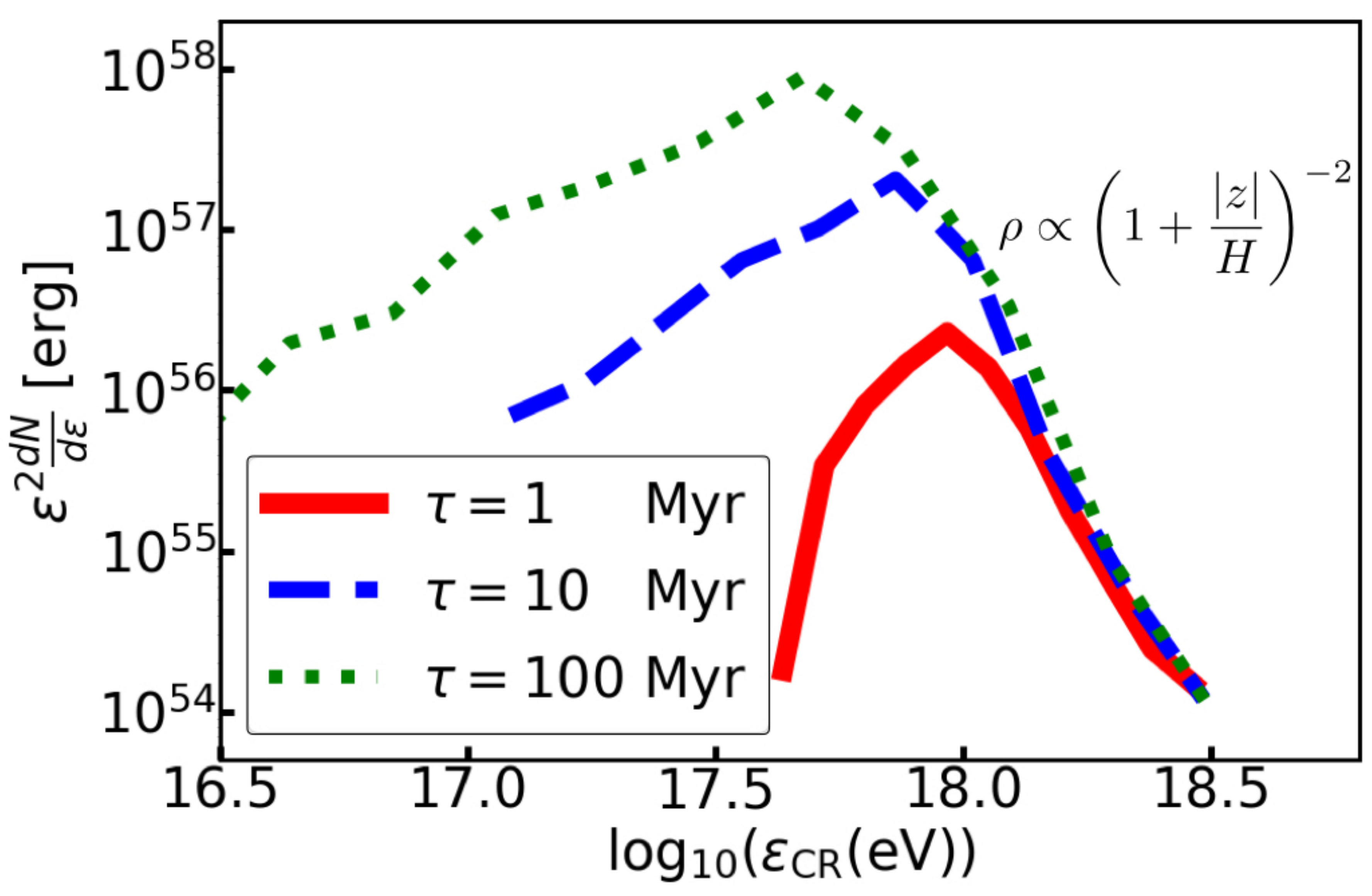}
    \includegraphics[width=0.45\textwidth]{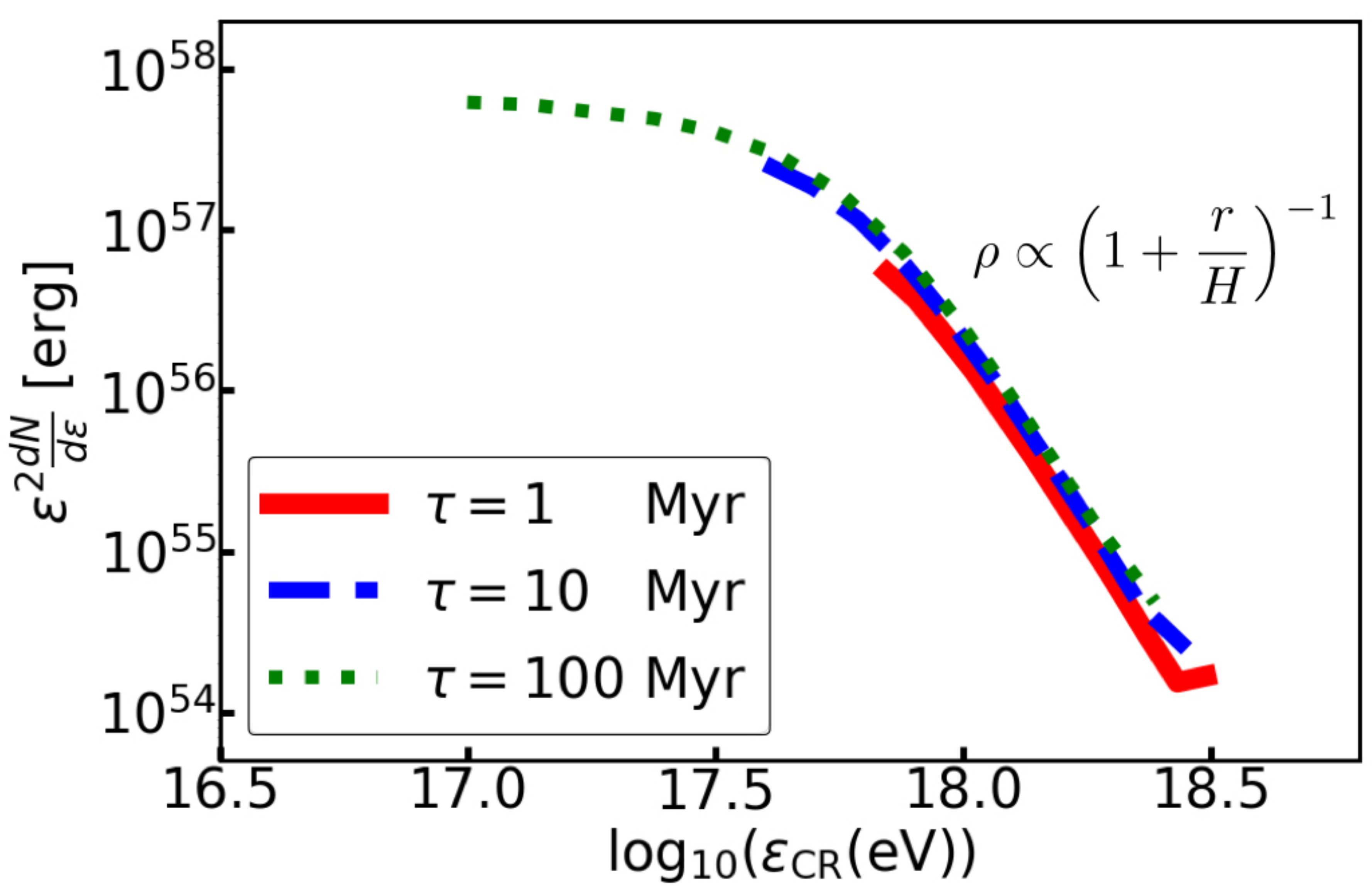}
    \includegraphics[width=0.45\textwidth]{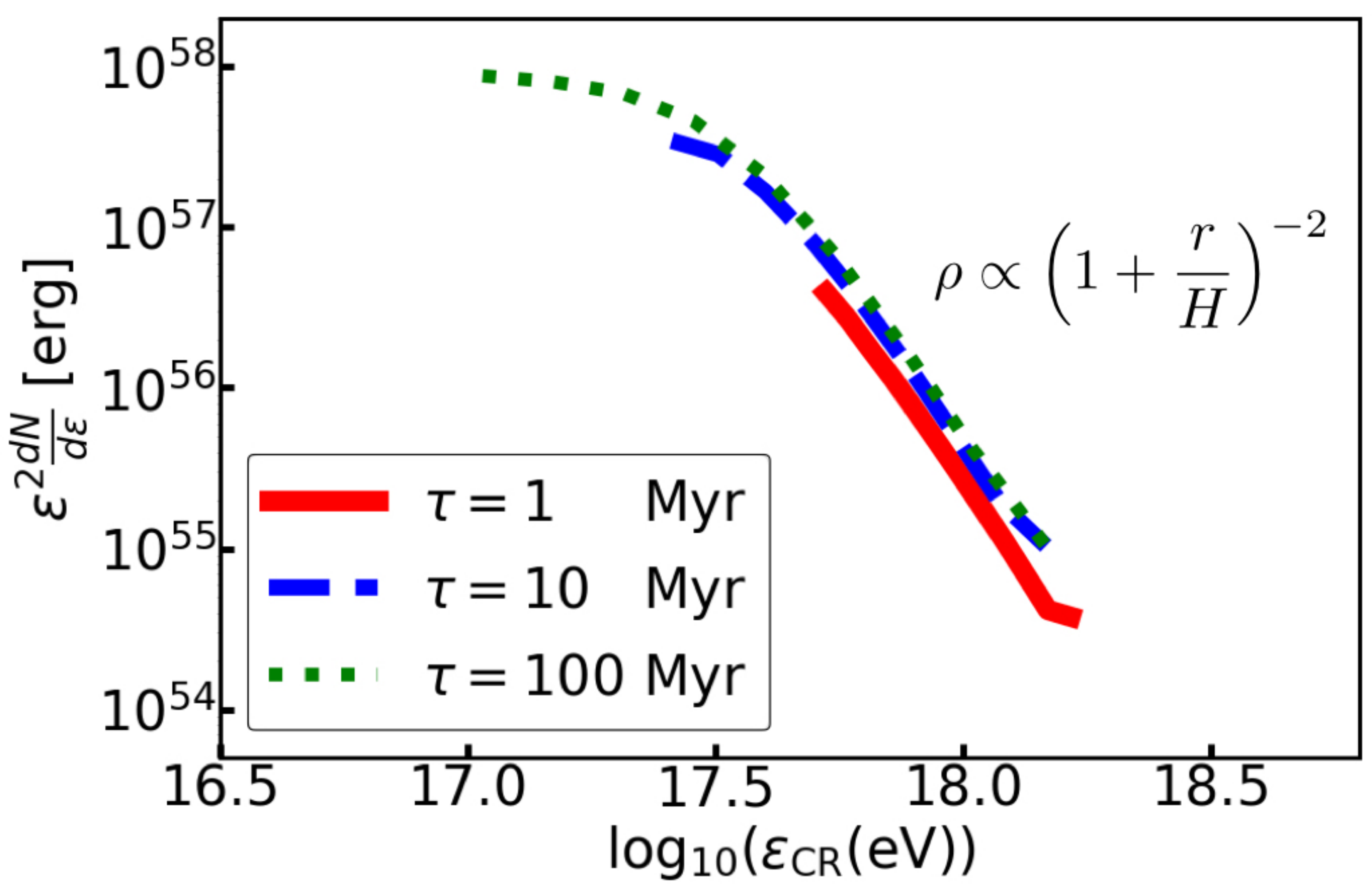}
    \caption{Plots of cumulative escaping CR spectra for four power-law density profiles: the top-left, top-right, bottom-left, bottom-right plots are vertically decaying with $\gamma = 1, 2$, and radially decaying with $\gamma = 1, 2$. }
    \label{fig:2}
\end{figure*}

\subsection{Escaping CR Spectra for Power-Law Density Profiles}

The constant and exponential density profiles are approximations to the gas distribution in a disc, 
while the power-law density profiles can serve as approximations to the gas distribution in a
Galactic gas halo \citep[e.g.][]{Kataoka2018}, where the gas distribution may extend out to a certain radius. For example, the radii are $\sim$10 kpc for several nearby spiral galaxies and out to between 18 kpc and 200 kpc for our Milky Way \citep[e.g.][]{Li2008, Yamasaki2009}. For gas haloes similar to that of the Milky Way,  gas distributions can be taaken to approximately follow power-law distributions that decrease spherically. But an exact form is hard to determine, so we are interested in both $\gamma = 1$ and $2$ cases, which resemble slow and fast decreasing scenarios. Meanwhile, it's also interesting to study gas haloes that decrease in a specific direction, which can be used as comparisons against the spherical distributions. Hence we also consider the power-law distributions that decay along one direction with $\gamma = 1, 2$.

We organize these four power-laws in a single subsection, for easier comparison. As the previous exponential case, the thermal luminosity needs to be fitted and energy conservation needs to be monitored throughout the calculations. In these four power-law cases the same conditions $\text{SFR}_{4} = 1, H = 1~\text{kpc}, E_{\text{ej}, 51} = 1$ are applied, and the central densities are normalised such that the gas haloes have the same total masses as that in the Galactic disc described by the previous exponential density profile. The total mass of the disc $M_{\text{disc}} = \int^{+\infty}_{-\infty}\pi R_{\text{disc}}^{2}\rho_{0, \text{disc}}\text{exp}(-z/H)\text{d}z$ is taken to be equal to $M_{\text{halo}} = \int^{R_{\text{halo}}}_{0}4\pi r^{2}\rho_{0, \text{halo}}/(1 + r/H)^{1, 2}\text{d}r$ to obtain $\rho_{0, \text{halo}}$, where $R_{\text{disc}}$ and $R_{\text{halo}}$ are assumed to be 10 kpc and 60 kpc, resembling the approximated disc radius and gaseous halo extension of the Milky Way.  The exact values can always be adjusted for systems of interest, and we normalise the total mass to the same value for different systems simply for more coherent comparisons. The cumulative escaping CR spectra are shown in Fig.~($\text{\ref{fig:2}}$). 

As shown by the Fig.~($\text{\ref{fig:2}}$), these four power-law cases resemble each other at early epochs, they all mimick initially the constant solutions, but they begin to behave differently at later ages. This is expected because at early times densities do not vary much for all cases, thus they resemble the constant density profile; while at later times the densities vary with respect to the density profiles, resulting in distinct spectra. After comparing the top two escaping CR spectra with the bottom two, we notice that for power-laws decaying along vertical directions, the spectra resemble the exponential case: the spectra peak at certain energies according to the time-scales up to which they are integrated. However, for the bottom two spherical cases, they resemble the constant density profile, in that they seem to to converge to the Sedov-Taylor scenario at later epochs gradually. The results also show that the top two cases can produce escaping CRs over more extensive energy ranges, while the bottom two have narrower ranges. This arises from the choice of cutoff points of the calculations. In this demonstration, the calculations are cut off according to time-scales which are fixed across different scenarios for coherent comparisons. However, if the calculations are integrated to a larger time-scale (physically still within the halo volume assumed), the top two cases would have broader spectra, while the bottom two continue to converge more obviously to the Sedov-Taylor scenario.

From another perspective, if grouped by indices, the results show that the $\gamma = 1$ cases can accelerate CRs to higher energies than the $\gamma = 2$ cases. At first glance, this is surprising because we expect that a lower ambient density is more favourable for higher energy CRs since the shock can propagate faster. This is caused by the normalisation used to derive the central density $\rho_{0}$ in the calculations. Since we normalise all of these density profiles to a fixed total gas mass, the central densities are lower in the $\gamma = 1$ cases than $\gamma = 2$ cases. Meanwhile, as shown from the plots, most of the higher energy escaping CRs are produced at early epochs when the shock hasn't moved far away from the centre. Thus the ambient gas densities can be approximated as the central gas densities. On the other hand, $\varepsilon_{\text{max}}$ can be shown to be $\propto \rho_{0}^{-1/10}$ at early epochs, thus a lower central density ($\gamma = 1$) can produce higher energy CRs than a high central density case ($\gamma = 2$). 
This comparison can be applied to all four cases since the total mass is normalised to the same value. It can be shown that $\rho_{0, \text{z}, \text{1}} < \rho_{0, \text{z}, \text{2}} < \rho_{0, \text{r}, \text{1}} < \rho_{0, \text{r}, \text{2}}$ (normalised central densities for vertical power-law cases with $\gamma = 1$ and 2, and spherical power-law cases with $\gamma = 1$ and 2), thus the maximal escaping energies for these four cases are reversely ordered, which is what is shown in Fig.~($\text{\ref{fig:2}}$). The jitters in the plots are an artifact due to the binning strategies used, which only lead to trivial changes to the results.

It is worth noting that for density profiles which quaalitatively resemble either a constant or an exponential profile, the escaping CR spectra can behave in an accordingly similar manner. For example, density profiles decaying along one direction (spherically) would produce a similar escaping CR spectrum as that of an exponential (constant) case does. Thus a complicated system can be simplified into a constant-like or exponential-like density scenario when studying the escaping CR spectra, since the results from the simplification, although not exactly accurate, can give a rough idea of the accurate underlining results. Of course, if a system is known to a certain extent, the most precise density profile should always be applied to perform the most accurate calculations. As an example, we will employ our model to a possible Galactic superbubble that originates from the centre of the Milky Way (MW). 

\section{Possible Explanation of the Observed Milky Way Cosmic-Ray Spectrum}
\label{sec:Application}

In this section, we apply our bubble model to a Galactic superbubble originating from the centre of the MW, assuming that the Galactic centre (GC) has been the host of a much stronger star-forming (SF) activity in the past than what is its current activity. For this simple application, we assume that the environment density profile follows a spherical hot gas halo distribution such as assumed for the MW by \cite{Kataoka2018}:
\begin{equation}
n(r) = n_{0}\left[1 + \left(\frac{r}{r_{\text{c}}}\right)^{2}\right]^{-3\beta/2}~,
\end{equation}
where $r$ is the distance to the GC, $n_{0}$ is the gas number density at $r=0$, fitted to be 0.46 $\text{cm}^{-3}$ by observations. The $r_{\rm c}$ is the core radius fitted to be 0.35 kpc and $\beta$ is $\simeq$ 0.71. 

We assume the radius (height) of the superbubble is $R = 9$ kpc and since $\beta\simeq 0.71$ and $3\beta/2$ is $\sim 1$, the density profile can be simplified to $n(r)\simeq n_{0}/(1+r/r_{\rm c})^{2}$ at large distance. Thus the spherical power-law density profile with $\gamma = 2$ should be applied, leading to the solution described in Section $\ref{sec:RadialPowerLaw2}$. With the power-law solution in Eq.~($\ref{eq:PL2RSolution}$), it can be shown that at $y \simeq 3.29~r_{\rm c}$ the height of the bubble is $\sim$ 9 kpc. Thus the evolution of the superbubble can be followed by performing calculations from $y = 0$ to $y\simeq3.29~r_{\rm c}$.  

\begin{table*}
\centering
\begin{tabular}{|c|c|c|c|c|c|c|}
\hline
      & SFR  $(\text{M}_{\odot}\text{yr}^{-1}
)$& $\tau_{\text{SFR}}$(\text{Myr}) & $T_{\text{age}}(\text{Myr})$  & $E_{\text{in}} (\times 10^{55} \text{erg})$                          & $E_{\rm k}(\times 10^{55} \text{erg})$      & $s_{p}$                    \\ \hline
Parameter Set 1 & 0.6 & 12.9     & 52.2 & 5.6 & 4.2  & 2.05  \\ \hline
Parameter Set 2 & 4.6  & 5.4     & 27.0 & 19.0 & 15.7   & 2.20   \\ \hline
Parameter Set 3 & 12.9  & 2.3     & 20.8 & 22.0 & 17.2   & 2.40   \\ \hline
\end{tabular}
\caption{Summary of the three parameter sets as well as the required proton acceleration spectrum indices, electron energy transfer efficiencies, and downstream magnetic fields considered in this paper. SFR is the star formation rate of the source, $\tau_{\text{SFR}}$ is the star-forming lifetime, $T_{\text{age}}$ is the calculated superbubble age, $E_{\text{in}}$ is the total input energy from the source, $E_{\rm k}$ is the total kinetic energy of the bubble shell (without taking account of energy losses and radiation), $s_{p}$ is the accelerated proton spectrum index.}
\label{tab:1}
\end{table*}

The upstream magnetic field of the superbubble at the current epoch is chosen to be $B_{\rm u}\sim 10~\mu$G, similar to that expected on the Fermi Bubbbles (FBs) \citep[e.g.][]{Guo2012a, Mou2015}. In this case, the actual values of $\epsilon_{B}$ are different from 0.01 as that in Section $\ref{sec:CREscapingSpectrum}$. For different parameter sets we calculate the dynamics first, then through Eq.~($\ref{eq:BFieldKineticE}$) with $B$ fixed as $10~\mu$G at the current epoch, we obtain the value of $\epsilon_{B}$. We then fix this efficiency $\epsilon_{B}$ and apply it in Eq.~($\ref{eq:BFieldKineticE}$) to calculate the magnetic field on the upstream at earlier epochs for each parameter set. Since there are essentially unlimited choices of the power and starburst time-scale of the source, we here adopt three specific parameter sets as demonstrations:  set 1 has a source of power $\text{SFR} = 0.6~ \text{M}_{\odot}\text{yr}^{-1}$ that lasts $\tau_{\text{SFR}} = 12.9~ \text{Myr}$, set 2 is of $\text{SFR} = 4.6~ \text{M}_{\odot}\text{yr}^{-1}$ that lasts $\tau_{\text{SFR}} = 5.4~ \text{Myr}$, and set 3 is of $\text{SFR} = 12.9~ \text{M}_{\odot}\text{yr}^{-1}$ that lasts $\tau_{\text{SFR}} = 2.3~ \text{Myr}$. These parameters are listed in Tab.~($\ref{tab:1}$). The obtained  $\epsilon_{B}$ are $7.9\%, 4.5\%$, and $2.8\%$, for set 1, set 2, and set 3, respectively. The reasons for choosing these three sets will be discussed in the following sections. 

\subsection{Results on the CR Flux and Spectrum Observed on the Earth}

For the parameter sets chosen, the escaping CRs can be obtained following the calculation scheme described in Section $\ref{sec:CREscapingSpectrum}$. The escaped CRs then undergo diffusive transport and the CR flux at the Earth can be calculated for each parameter set. 
For a spherical halo size of $R_{\rm h}$, the diffusion time-scale, $T_{\text{diff}} = R_{\rm h}^{2}/6D_{\text{h}}(\varepsilon)$, is compared with the ages of the superbubble to determine whether this system can be treated as a bursting injection, where $T_{\text{diff}} > T_{\text{age}}$, or as a continuous injection, when $T_{\text{diff}} < T_{\text{age}}$. In this section, since this quantity is uncertain, the CR halo size is assumed to be 10 kpc \citep[e.g.][]{Delahaye2010}, slightly larger than the scale of the superbubble. Since we expect high-energy CRs to be produced by the superbubble, only the CRs around and above the knee energy, $\varepsilon_{\text{knee}} = 10^{15.5}$ eV, are considered. For the bursting injection, one could use the solution shown in Appendix A of \cite{Fujita2017}. However, this does not seem to be the appropriate case, because at and above the knee energy $T_{\text{diff}}\leq~1.1\text{Myr}\left(R_{\rm h}/10\text{kpc}\right)^{2}\left(D_{\text{h}, \text{knee}}/4.4\times10^{30}\text{cm}^{2}\text{s}^{-1}\right)^{-1}$, which is less than the three bubble ages considered in this section.  

Thus, we focus on the continuous injection case, $T_{\text{diff}} < T_{\text{age}}$. Then the CR spectrum is estimated by
\begin{equation}
E^{2}\Phi\simeq \frac{(EL_{E, \text{inj}})X_{\text{esc}}}{4\pi M_{\text{gas}}}~,
\label{eq:ContinuousInjectionCRSpectrum}
\end{equation}
where $EL_{E, \text{inj}}$ is the CR energy injection luminosity obtained from applying our model, $M_{\text{gas}}$ is the total gas mass contained inside the CR halo, estimated to be $\simeq 1.0\times10^{10}~\text{M}_{\odot}$, and $X_{\text{esc}}$ is the grammage along the CR path length, which is obtained via observations on the ratio of boron to carbon fluxes \citep[e.g.][]{Adriani2014, Aguilar2016}. The above approximation is valid as long as the grammage is dominated by the gas mass in the disc region~\citep{MF19}. 

\begin{figure}
    \centering
    \includegraphics[width=0.46\textwidth]{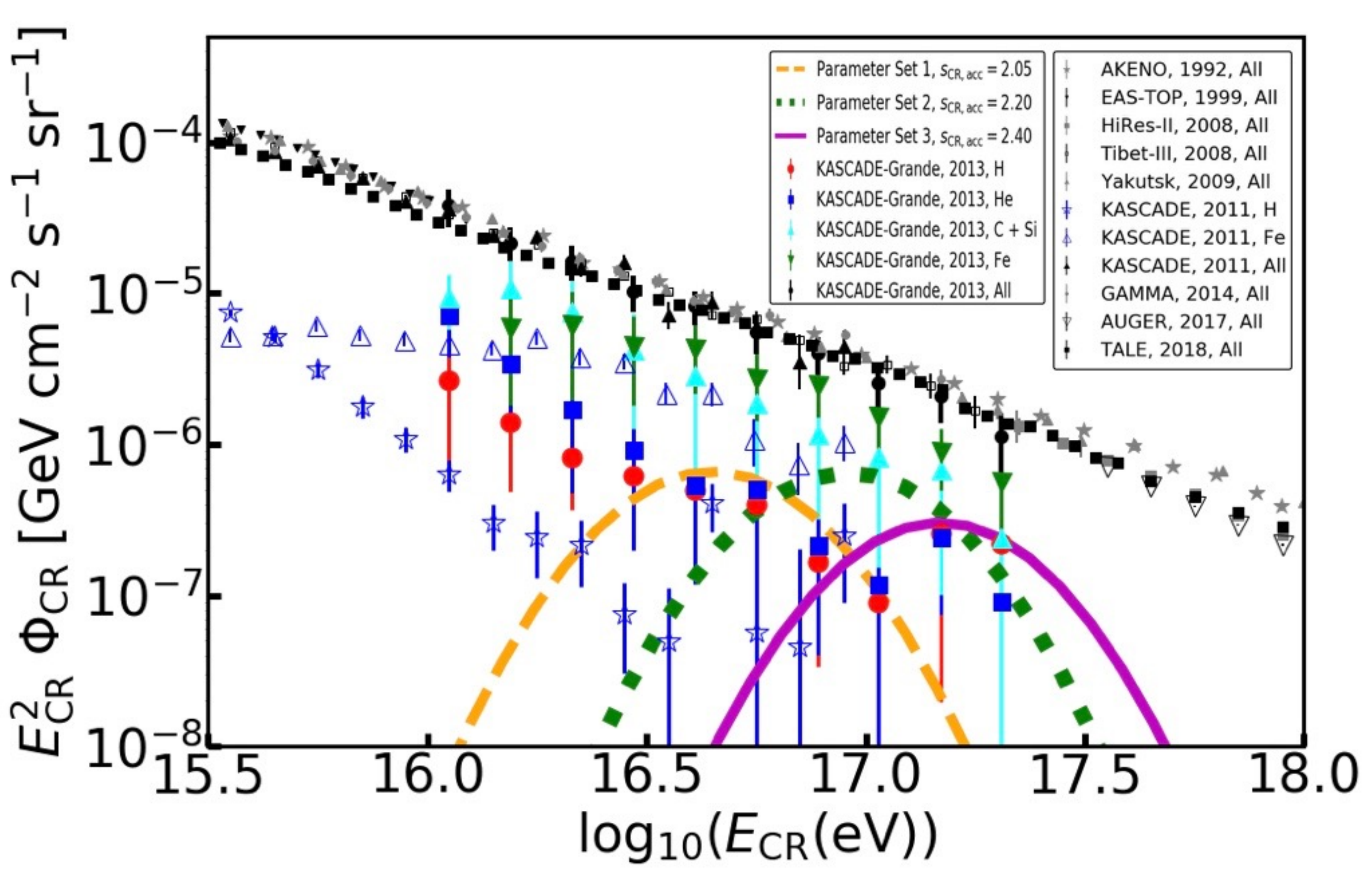}
    \caption{CR fluxes calculated from three parameter sets. The black and grey data are the observed CR overall spectrum, the orange-dashed line is the CR flux calculated for parameter set 1, the green-dotted line is that for parameter set 2, and the purple-solid line is that for parameter set 3. The red, round and blue, square data points are hydrogen and helium nuclei components of CRs above the knee from KASCADE-Grande \citep[][]{Apel2013}.  Rest of the shown data are taken from AKENO, 1992, All: \citet{Nagano1992}; EAS-TOP, 1999, All: \citet{Aglietta1999}; HiRes-II, 2008, All: \citet{Abbasi2008}; Tibet-III, 2008, All: \citet{Amenomori2008}; Yakutsk, 2009, All: \citet{Ivanov2009}; KASCADE, 2011: \citet{Finger2011}; GAMMA, 2014, All: \citet{Ter-Antonyan2014}; AUGER, 2017, All: \citet{Fenu2017}; TALE, 2018, All: \citet{Abbasi2018}.}
    \label{fig:3}
\end{figure}

In previous sections $\epsilon_{p} = 0.1$ is used to demonstrate the generic properties of the escaping CR spectra, but simulations have shown that the efficiency can vary \citep[e.g.][]{Caprioli2014}. To be conservative,  $\epsilon_{p} = 0.01$ is used in this section, the reason for choosing this value being clarified in the subsequent discussion (note that we consider only protons here, instead all of the CRs). In the general discussions (Section $\ref{sec:CREscapingSpectrum}$), a fixed $s_{p} = 2$ acceleration CR spectrum index is used, but here we treat this value as a variable with a range of [2.0, 2.4].  Thus by varying the spectral index, the escaping efficiency can be changed (a combined effect of acceleration efficiency and spectrum index), specific values being determined depending on different purposes. The results are shown in Fig.~($\ref{fig:3}$). In the plot, the orange-dashed line is the CR flux calculated for the parameter set 1, the green-dotted line is that for the parameter set 2, and the purple-solid line that for the set 3. The red, round and blue, square data points are proton and helium nuclei components of CRs measured above the knee by the KASCADE-Grande \citep[e.g.][]{Apel2013}. The unfilled blue data points are proton and iron components measured by the KASCADE \citep[e.g.][]{Finger2011}. The black and grey data points are measurements of all particle CRs by different experiments. 

As shown by the figure, the parameter set 1 is able to reproduce well the decrease of the H and He CR components from $\sim 10^{16.5}$ eV to $\sim 10^{17}$ eV with index $s_{p} = 2.05$. The conventional theory considers that the first knee of the CR spectrum is the result of Galactic SNe which accelerate different nuclei to different maximum energies that are proportional to the atomic numbers of the nuclei. Thus, light elements, such as H and He, would not be able to obtain energies much above the knee, leading to the decline of the light-element abundance after the knee. However, even though SNR are believed to be able to accelerate protons up to a few PeV, they have not been observed to be PeV accelerators of CR protons (so-called `Pevatrons') \citep[e.g.][]{Fujita2017}. Our model (parameter set 1) demonstrates that the SF activity that produced our possible Galactic superbubble can provide a potential alternative for accelerating light elements above the knee, which is consistent with the argument of \cite{MF19} (see their Eq.~29). 
This is the reason why we choose the parameter set 1 as a specific case to study here. 

The KASCADE-Grande observations show that there is an increase of H and He abundance in the CRs after $\geq10^{17}$ eV, which is puzzling, and many models have been proposed to solve it \citep[e.g.][]{Apel2013}. It is the general belief that CRs above the second knee are produced extragalactically \citep[e.g.][]{Aloisio2007, Kampert2012, Pierre2017}. Our bubble model (with the parameter set 3), indicated by the purple-solid line, demonstrates that the source that produced our Galactic superbubble can generate also such an increase of light elements at these energies in the observed CR flux. The spectral index is $s_{p} = 2.40$, the steepest value in the range. The parameter set 2 is an intermediate case used to demonstrate that it is possible for a single parameter set to produce both the light CRs below and above $10^{17}$ eV. The spectral index in this case is $s_{p} = 2.20$. 

\subsection{Discussion}
\label{sec:Discussion}

In this section, the proton acceleration efficiency ($\epsilon_{p}$) was fixed at 0.01 when performing the calculations. As shown by Fig.~($\ref{fig:3}$), with a spectral index $s_{p} = 2.4$ and $\epsilon_{p} = 0.01$, the calculated CR flux can reproduce the rise in the detected light elements above the second knee without overshooting the observations. To make the entire calculation consistent, $\epsilon_{p}$ is fixed and the spectral index is varying. In the same plot, the required spectral index for the parameter set 1 is $s_{p} = 2.0$ at $\epsilon_{p} = 0.01$. However, since the escaping CR spectra do not change much for slightly different parameters (as discussed in Section $\ref{sec:CREscapingSpectrum}$), a similar result for parameter set 1 can be obtained by using a different set of $s_{p}$ and $\epsilon_{p}$, for example $s_{p} = 2.4$ and $\epsilon_{p} = 0.1$. Thus there exists a degeneracy in the choice of variables for the parameter set 1 (and similarly for set 2). This degeneracy does not exist for the parameter set 3 because either an increase in the acceleration efficiency or a decrease of the spectral index will boost the CR flux, overshooting the observations. 

The CR halo size is assumed to be greater than the size of our superbubble when Eq.~($\ref{eq:ContinuousInjectionCRSpectrum}$) is used. However, given the observational uncertainties, the CR halo size could be as large as 15 kpc, or as small as 0.3 kpc \citep[e.g.][]{Protheroe1982, Moskalenko1998, Delahaye2010, Blum2013}. If the CR halo size is less than the size of the superbubble, a fraction of produced CRs will escape freely from the halo, leading to a reduction in the  calculated CR flux at the Earth. This can be estimated from the ratio of the time-integrated area within CR halo size to the total time-integrated area of the superbubble. For example, a halo size of 0.65 kpc causes a reduction of the calculated CR flux by $\sim90$ per cent. Hence $\epsilon_{p} \sim 0.1$ is needed to obtain the same level of observed CR flux. 

%
The main purpose of this work is to calculate the flux of the escaping CRs. 
The confined CRs that remain in the bubble carry a larger amount of energy than that of the escaping ones, although 
the individual CRs themselves have lower energies than those of the observed escaping CRs. 
If the trapped CRs eventually diffuse out from the superbubble, they would become observable. 
However, at the energies we are interested in, $\sim 10^{17}$ eV, we find that the CR flux is dominated by escaping CRs. 
An approximate calculation shows that the flux of trapped CRs, if they escape without energy losses, is 
$\sim 10^{-8}~\text{GeV}~\text{s}^{-1}~\text{cm}^{-2}~\text{sr}^{-1}$, 
which is less than the observed escaping CR flux at this energy level. 
In reality, they are also subjected to adiabatic losses due to the expansion of the superbubble before they escape. 
Furthermore, we note that the CRs need to leave the superbubble to contribute to the CR flux observed on the Earth. 
The nominal size of the bubble used in our calculation is 10 kpc, implying that the bulk portion of the bubble 
is above the galactic disc. 
Although we do not calculate the bubble propagation in the galactic disc, 
the gas distribution in the disc has a much larger density than in the gas halo, and the
disc gas pressure has a much greater resistance for the bubble propagation along the horizontal directions.
Thus, given for a roughly constant distribution of the gas in the galactic disc, through Eq. $\ref{eq:RsConstant}$
it can be shown that the bubble would not advance more than about 4 kpc from the GC, i.e. it stops far 
from the Earth.

Our calculation about the superbubble sets a stage for an application to the Fermi Bubbles (FB), a bubble-like, bi-lobular structure extending out to $\sim$ 9 kpc symmetrically around the GC, which is one of the most important discoveries by the $Fermi$-LAT instrument \citep[e.g.][]{Su2010}.  The FB has a sharp edge at $\sim10~$GeV gamma-rays and produces almost uniform brightness across all the surface \citep[e.g.][]{Atwood2009, Kataoka2018}. Because of the symmetric structure of the FB, activities related to the GC are naturally preferred explanations, for example, stellar winds, starburst activity, supernovae, jets, and so on \citep[e.g.][]{Su2010, Crocker2011, Guo2012a, Guo2012b, Lacki2014, Ackermann2014}, but the observations have not been able to pinpoint the exact model yet~\citep[e.g.,][]{Ahlers2014}. The exact application of our model to the FB is beyond the scope of the present paper, but it remains as an interesting object to be studied in the future. 

Besides the FB, our calculation scheme has a wider range of applicability. For example, it could 
be applied to superbubbles in various starburst galaxies, such as Arp 220 \citep[e.g.][]{Anantharamaiah2000, Paggi2017}
and extragalactic `Fermi Bubbles' in NGC 3079 \citep[e.g.][]{Li2019}. 
Furthermore, by convolving the CR  production rate in individual superbubbles with the star formation rate function 
we can calculate the cumulative CRs from starburst galaxies by integrating over a broad range of SFR.  
We also note that one of the consequences of such models is that neutrinos and gamma-rays may be expected from $pp$ interactions produced by the escaping CRs into the surrounding medium~\citep[e.g.][]{Ahlers2014,Senno2015, Xiao2016}.

\section{Summary}
\label{sec:Summary}

In this paper, we explored the consequences of the fact that the extreme star formation rate in starburst galaxies can naturally produce superbubbles. The propagation of the resulting shocks in the interstellar and circumgalactic medium  has been one of the principal areas of study, since understanding their properties directly links the observations to the underlying physics. Analytical solutions usually exist for spherical systems, but it is also of great interest to study shocks propagating into nonuniform/nonisotropic ambient environments, for example, the atmosphere of the Earth. Kompaneets' shock propagation solution is regarded as the first 2-D solution in a stratified exponentially decreasing density profile \citep[][]{Kom1960}, and since then, many efforts have been made to study shock-propagation in nonuniform systems, but only a few special cases have analytic solutions \citep[for a review, see][]{Bisnovatyi1995}. An interstellar bubble powered by a strong wind was first studied by \cite{Castor1975} under a constant density profile. Superbubbles, which might be powered by semi-continuously produced SNe, can be treated like very large wind bubbles \citep[e.g.][]{MacLow1988}, hence a similar mechanism can be applied. We showed that ions could be accelerated up to super-knee energies by such a superbubble through diffusive shock acceleration.

For non-uniform environments, however, the solutions given by the constant case are no longer valid. We applied the Kompaneets' formalism to obtain the analytic solutions for bubbles propagating in different density profiles, which were then used to calculate the spectra of the escaping CRs from the bubble shell using the `escape-limited' model of \cite{Ohira2010} in an approximate manner. This model with a free escape boundary has been used in many scenarios \citep[e.g.][]{Zhang2018, Ohira2018}, but this is the first time that it is used in combination with a model for a superbubble propagation into a non-uniform density. We found that a similar $\text{d}N/\text{d}\varepsilon\propto \varepsilon^{-6.89}$ spectrum is produced at early times for different cases. This can be explained approximately as a combination of a series of Sedov-Taylor type explosive inputs from frequent SNe. At later times, after the starburst activity has ceased and the bubble has moved far away from the centre, the spectra can behave very differently.

We applied our model to a possible Galactic superbubble in Section $\ref{sec:Application}$, for three specific parameter sets of energy input, listed in Tab.~($\ref{tab:1}$), and we calculated the CR flux arriving at the Earth resulting
from these three parameter sets. We found that the parameter set 1 can naturally reproduce the observed decline of the 
light-element (H + He) abundances below $10^{17}$ eV, while the parameter set 3 is able to explain the observed rise of the light-element components around and above $10^{17}$ eV, while the parameter set 2 serves as an intermediate scenario.   

We expect that the model developed here can also be applied to extragalactic superbubbles, i.e., starburst activities in galaxies other than the Milky Way.

\section*{Acknowledgements}
We acknowledge partial support from NASA NNX13AH50G (ZZ and PM), the Alfred P. Sloan Foundation, NSF grant No. PHY-1620777 and No. AST-1908689 (KM). 
We are grateful to Yu Jiang, Shigeo Kimura, and Chengchao Yuan for useful discussions. 
We also thank Donald Elison for reviewing this manuscript with useful comments and suggestions.

\bsp	
\label{lastpage}

\begin{thebibliography}{99}

\bibitem[\protect\citeauthoryear{Abbasi et al.}{2008}]{Abbasi2008} Abbasi R. U. et al.,\ 2008, \prl, 100, 101101

\bibitem[\protect\citeauthoryear{Abbasi et al.}{2018}]{Abbasi2018} Abbasi R. U. et al.,\ 2018, \apj, 865, 1

\bibitem[\protect\citeauthoryear{Ackermann et al.}{2014}]{Ackermann2014}Ackermann M. et al.,\ 2014, \apj, 793, 64

\bibitem[\protect\citeauthoryear{Adriani et al.}{2014}]{Adriani2014} Adriani O. et al.,\ 2014, \apj, 791, 93

\bibitem[\protect\citeauthoryear{Aglietta et al.}{1999}]{Aglietta1999} Aglietta M. et al.,\ 1999, Astropart. Phys. 10, 1

\bibitem[\protect\citeauthoryear{Aguilar et al.}{2016}]{Aguilar2016} Aguilar M. et al.\ 2016, \prl, 117, 231102

\bibitem[\protect\citeauthoryear{Ahlers \& Murase}{2014}]{Ahlers2014} Ahlers M., Murase K.,\ 2014, \prd, 90, 023010

\bibitem[\protect\citeauthoryear{Aloisio et al.}{2007}]{Aloisio2007} Aloisio R. et al.,\ 2007, Astropart. Phys., 27, 76

\bibitem[\protect\citeauthoryear{Amenomori et al.}{2008}]{Amenomori2008} Amenomori M. et al.,\ 2008, \apj, 678, 2

\bibitem[\protect\citeauthoryear{Anantharamaiah et al.}{2000}]{Anantharamaiah2000} Anantharamaiah K. R. et al.,\ 2000, \apj, 537, 613

\bibitem[\protect\citeauthoryear{Apel et al.}{2013}]{Apel2013} Apel W. D. et al.,\ 2013, Astropart. Phys., 43, 71

\bibitem[\protect\citeauthoryear{Atwood et al.}{2009}]{Atwood2009} Atwood W. B. et al.,\ 2009, \apj, 697, 2 

\bibitem[\protect\citeauthoryear{Axford, Leer, \& Skadron}{1997}]{Axford1977} Axford W. I., Leer E., Skadron G.,\ 1977, Proc. 15th Int. Cosmic Ray Conf., Plovdiv, 11, 132

\bibitem[\protect\citeauthoryear{Basu, Johnstone, \& Martin}{1999}]{Basu1999} Basu S., Johnstone D., Martin P.~G.,\ 1999, \apj, 516, 843
 
\bibitem[\protect\citeauthoryear{Bell}{1978}]{Bell1978} Bell A. R.,\ 1978, \mnras, 182, 147

\bibitem[\protect\citeauthoryear{Blandford \& Eichler}{1987}]{Blandford1987} Blandford R. D., Eichler D.,\ 1987, \physrep, 154,1

\bibitem[\protect\citeauthoryear{Blum, Katz, \& Waxman}{2013}]{Blum2013} Blum K., Katz B., Waxman E.,\ 2013, \prl, 111, 211101

\bibitem[\protect\citeauthoryear{Bisnovatyi-Kogan \& Silich}{1995}]{Bisnovatyi1995}  Bisnovatyi-Kogan G. S., Silich S. A., \ 1995, Rev. Mod. Phys. 67, 661

\bibitem[\protect\citeauthoryear{Caprioli, Amato, \& Blasi}{2010}]{Caprioli2010} Caprioli D., Amato E., Blasi P.,\ 2010, Astropart. Phys. 33, 160

\bibitem[\protect\citeauthoryear{Caprioli \& Spitkovsky}{2014}]{Caprioli2014} Caprioli D., Spitkovsky A.,\ 2014, \apj, 783, 91

\bibitem[\protect\citeauthoryear{Castor, McCray, \& Weaver}{1975}]{Castor1975} Castor J.,  McCray R., Weaver R.,\ 1975, \apj, 200, 107

\bibitem[\protect\citeauthoryear{Crocker \& Aharonian}{2011}]{Crocker2011} Crocker R. M., \& Aharonian F., \ 2011, \prl, 106, 101102

\bibitem[\protect\citeauthoryear{Delahaye et al.}{2010}]{Delahaye2010} Delahaye T. et al.,\ 2010, \aap, 524, A51

\bibitem[\protect\citeauthoryear{Drury}{1983}]{Drury1983} Drury L.,\ 1983, Rep. Prog. Phys., 46, 973

\bibitem[\protect\citeauthoryear{Drury}{2011}]{Drury2011} Drury L.,\ 2011, \mnras, 415, 1807

\bibitem[\protect\citeauthoryear{Fenu et al.}{2017}]{Fenu2017} Fenu F. for the Pierre Auger Collaboration, \ 2017, PoS ICRC2017 (2018) 486

\bibitem[\protect\citeauthoryear{Finger}{2011}]{Finger2011} Finger M.,\ 2011, Ph.D. thesis. http://d-nb.info/1014279917/34

\bibitem[\protect\citeauthoryear{Fujita, Murase, \& Kimura}{2017}]{Fujita2017} Fujita Y., Murase K., Kimura S.,\ 2017, JCAP04 037

\bibitem[\protect\citeauthoryear{Guo \& Mathews}{2012}]{Guo2012a} Guo F., Mathews W., \ 2012, \apj, 756, 181

\bibitem[\protect\citeauthoryear{Guo et al.}{2012}]{Guo2012b} Guo F., Mathews W., Dobler G., Peng OH S.,\ 2012, \apj, 756,182

\bibitem[\protect\citeauthoryear{Ivanov, Knurenko, \& Sleptsov}{2009}]{Ivanov2009} Ivanov A. A., Knurenko S. P., Sleptsov I., Ye.,\ 2009, New Journal of Physics, 11, 065008

\bibitem[\protect\citeauthoryear{Kampert \& Unger}{2012}]{Kampert2012} Kampert K.-H., Unger M.,\ 2012, Astropart. Phys., 35, 660

\bibitem[\protect\citeauthoryear{Kataoka et al.}{2018}]{Kataoka2018} Kataoka J., Sofue Y., Inoue Y., et al.,\ 2018, Galaxies, 6(1), 27

\bibitem[\protect\citeauthoryear{Krymsky}{1977}]{Krymsky1977} Krymsky G. F.,\ 1977, Doki. Akad. Nauk SSSR, 234, 1306

\bibitem[\protect\citeauthoryear{Kompaneets}{1960}]{Kom1960} Kompaneets A.~S.,\ 1960, Dokl. Akad. Nauk. SSSR, 130, 1001 [Soviet Phys. Dokl, 5, 46 (1960)]

\bibitem[\protect\citeauthoryear{Lacki}{2014}]{Lacki2014} Lacki B. C., \ 2014, MNRASL,  444, L39

\bibitem[\protect\citeauthoryear{Li et al.}{2008}]{Li2008} Li J. T. et al.,\ 2008, \mnras, 2008, 390, 59

\bibitem[\protect\citeauthoryear{Li et al.}{2019}]{Li2019} Li J. T. et al.,\ 2019, arXiv: 1901.10536

\bibitem[\protect\citeauthoryear{Mac Low \& McCracy}{1988}]{MacLow1988} Mac Low M., McCray R.,\ 1988, \apj, 324, 776

\bibitem[\protect\citeauthoryear{Mannucci et al.}{2010}]{Mannucci2010} Mannucci F. et al.,\ 2010, \mnras, 408, 2115


\bibitem[\protect\citeauthoryear{Moskalenko \& Strong}{1998}]{Moskalenko1998} Moskalenko I., Strong A.,\ 1998, \apj, 493, 694

\bibitem[\protect\citeauthoryear{Mou et al.}{2015}]{Mou2015} Mou G., Yuan F., Guan Z., Sun M., \ 2015, \apj, 811, 37

\bibitem[\protect\citeauthoryear{Murase \& Fukugita}{2019}]{MF19} Murase K., Fukugita M., \ 2019, \prd, 99, 063012

\bibitem[\protect\citeauthoryear{Nagano et al.}{1992}]{Nagano1992} Nagano M. et al.,\ 1992, Journal of Physics. G, Nuclear and Particle Physics (United Kingdom), 18, 2

\bibitem[\protect\citeauthoryear{Olano}{2009}]{Olano2009} Olano C.~A.,\ 2009, \aap, 506, 1215

\bibitem[\protect\citeauthoryear{Ohira, Murase, \& Yamazaki}{2010}]{Ohira2010} Ohira Y., Murase K., Yamazaki R.,\ 2010, \aap , 513, A17

\bibitem[\protect\citeauthoryear{Ohira, Kisaka, \& Yamazaki}{2018}]{Ohira2018} Ohira Y., Kisaka S., Yamazaki R.,\ 2018, \mnras, 478, 926

\bibitem[\protect\citeauthoryear{Paggi et al.}{2017}]{Paggi2017} Paggi A. et al.,\ 2017, \apj, 841, 44

\bibitem[\protect\citeauthoryear{Protheroe}{1982}]{Protheroe1982} Protheroe, R.,\ 1982, \apj, 254, 391

\bibitem[\protect\citeauthoryear{The Pierre Auger Collaboration}{2017}]{Pierre2017} The Pierre Auger Collaboration, \ 2007, Science 357 (6357), 1266

\bibitem[\protect\citeauthoryear{Rowan-Robinson}{2009}]{Rowan2009} Rowan-Robinson M.,\ 2009, \mnras, 394, 117

\bibitem[\protect\citeauthoryear{Rowan-Robinson \& Wang}{2010}]{Rowan2010} Rowan-Robinson M., Wang L.,\ 2010, \mnras, 406, 720

\bibitem[\protect\citeauthoryear{Rowan-Robinson et al.}{2017}]{Rowan2017} Rowan-Robinson M. et al.,\ 2017, \aap, 619, A169

\bibitem[\protect\citeauthoryear{Sanders \& Mirabel}{1996}]{Sanders1996} Sanders D. B., Mirabel F., \ 1996, \araa, 34, 794

\bibitem[\protect\citeauthoryear{Senno et al.}{2015}]{Senno2015} Senno N. et al., \ 2015, \apj, 806, 24

\bibitem[\protect\citeauthoryear{Soifer}{1984}]{Soifer1984} Soifer B.~T. et al.,\ 1984, \apj, 283, L1

\bibitem[\protect\citeauthoryear{Su, Slatyer, \& Finkbeiner}{2010}]{Su2010} Su M., Slatyer T. R., Finkbeiner D. P.,\ 2010, \apj, 724, 1044

\bibitem[\protect\citeauthoryear{Tacconi et al.}{2013}]{Tacconi2013} Tacconi L. J. et al.,\ 2013, \apj, 768, 74

\bibitem[\protect\citeauthoryear{Ter-Antonyan}{2014}]{Ter-Antonyan2014} Ter-Antonyan S.,\ 2014, \prd, 89, 123003

\bibitem[\protect\citeauthoryear{Xiao et al.}{2016}]{Xiao2016} Xiao, D., M\'{e}sz\'{a}ros P., Murase K., Dai Z.,\ 2016, \apj, 826, 133

\bibitem[\protect\citeauthoryear{Yamasaki et al.}{2009}]{Yamasaki2009} Yamasaki N. Y. et al.,\ 2009, \pasj, 69, S291


\bibitem[\protect\citeauthoryear{Zeldovich \& Raizer}{1966}]{Zeldovich1966} Zel'dovich Ya., B., Raizer Yu. P.,\ 1966, Physics of Shock Waves and High-Temperature Hydrodynamics Phenomena, Dover Publications, Mineola, New York

\bibitem[\protect\citeauthoryear{Zhang et al.}{2018}]{Zhang2018} Zhang B., Murase K., Kimura S., et al.,\ 2018, \prd, 97, 8

\end{thebibliography}
\end{document}